\documentclass{Interspeech2024}

\interspeechcameraready

\usepackage{amsmath}
\usepackage{amssymb}
\usepackage{mathtools}
\usepackage{amsthm}
\usepackage{booktabs}  
\usepackage{tabularx}  
\usepackage{caption} 
\newcommand{\bc}{\boldsymbol{c}}
\newcommand{\br}{\boldsymbol{r}}
\newcommand{\bz}{\boldsymbol{z}}
\newcommand{\bd}{\boldsymbol{d}}
\newcommand{\bp}{\boldsymbol{p}}
\newcommand{\bx}{\boldsymbol{x}}
\newcommand{\bo}{\boldsymbol{o}}
\newcommand{\bh}{\boldsymbol{h}}
\usepackage{svg}
\title{LiveSpeech: Low-Latency Zero-shot Text-to-Speech via Autoregressive Modeling of Audio Discrete Codes}

\name[]{Trung}{Dang}
\name[]{David}{Aponte}
\name[]{Dung}{Tran}
\name[]{Kazuhito}{Koishida}

\address{
  Applied Sciences Group, Microsoft Corp, USA}
\email{trungdang@, davidaponte@, dung.tran@, kazukoi@microsoft.com}


\begin{document}

\maketitle

\begin{abstract}
    
    Prior works have demonstrated zero-shot text-to-speech by using a generative language model on audio tokens obtained via a neural audio codec. It is still challenging, however, to adapt them to low-latency scenarios. In this paper, we present LiveSpeech - a fully autoregressive language model-based approach for zero-shot text-to-speech, enabling low-latency streaming of the output audio. To allow multiple token prediction within a single decoding step, we propose (1) using adaptive codebook loss weights that consider codebook contribution in each frame and focus on hard instances, and (2) grouping codebooks and processing groups in parallel. Experiments show our proposed models achieve competitive results to state-of-the-art baselines in terms of content accuracy, speaker similarity, audio quality, and inference speed while being suitable for low-latency streaming applications.
\end{abstract}

\section{Introduction}
\label{sec:intro}
Zero-shot text-to-speech (TTS) has gained attention in recent years due to its ability to synthesize speech that is similar to any voice without speaker-specific model adaptation \cite{vall-e,vall-ex,musicgen,naturalspeech2}. Although recent research in zero-shot TTS has made significant progress in achieving high audio quality and speaker similarity through the utilization of language models applied to tokenized audio \cite{vall-e,vall-ex} or diffusion models \cite{naturalspeech2}, there remains a challenge in adapting them to a real-time or low-latency setting. This challenge arises due to the non-autoregressive nature of some models or the high inference time per step associated with others. The development of a low-latency zero-shot TTS system holds the potential to unlock a diverse range of applications, particularly in facilitating live communication scenarios, including speech-to-speech translation, accent conversion, speech simplification, or disfluency removal.

In streaming applications, autoregressive models offer a distinct advantage due to their ability to generate speech incrementally, making them well-suited for tasks that require immediate responses. Recent research has demonstrated that zero-shot TTS can be accomplished by harnessing the capabilities of language models on discrete tokens obtained from neural audio codecs \cite{vall-e,audiolm}. However, due to the high bandwidth nature of audio, a single audio frame is usually represented by multiple codes, which may also be sequentially dependent \cite{soundstream} thus need to be predicted in sequential transformer steps. To speed up the generation, a delayed generation pattern \cite{musicgen} has been proposed to shift codes in each frame in order to produce codes from different frames within a single step, while codes from the same frame are produced in sequential steps. However, the bandwidth of one decoding step may limit the number of codes that can be predicted in parallel, since the model must maintain and process the information of all codes throughout its layers. While this can parallelize 4-codebook generation in the music generation task with a large model size \cite{musicgen},
it may not perform well in the low-latency TTS task for a lower model capacity, and a higher number of codebooks (e.g., 8 or 16) that are required to represent a wide range of subtle variations in human speech.

In this work, we propose LiveSpeech - a fully autoregressive transformer architecture for the zero-shot TTS task and demonstrate its competitive performance to existing approaches, as well as its ability to perform low-latency inference in a streaming manner. %
Our contributions can be summarized as follows: (1) We introduce a loss weighing mechanism to redistribute the model capacity across codebooks. We weigh each code based on its contribution to the constructed frame and whether more important codes in the same frame are accurately predicted with high confidence. Our model can efficiently scale the number of codebooks for each generated frame to 16 without additional inference cost, (2) we show that how enhancing the step capability by modeling groups of codebooks in parallel can further improve the performance. While the computation increases, codes in these groups can be predicted in parallel without introducing significant inference time.

\section{Related Works}
\label{sec:related_works}

Traditional works on speech generation adopt a transformer architecture to generate downsampled speech frames of mel-spectrograms \cite{tacotron2,ping2017deep,wu2022adaspeech,casanova2021sc}, which can be decoded to the raw audio by using a vocoder. However, generating mel-spectrogram is hard - it is susceptible to decoding noise and performs poorly on zero-shot or noisy condition \cite{mqtts}. Recently, the sequential and continuous nature of speech has provided inspiration for leveraging successful techniques employed in text generation, such as language models, and in image generation, such as diffusion models. Both the language modeling and diffusion approach have demonstrated their efficiency in generating high-fidelity audio in a wide variety of audio and speech generation tasks \cite{audiolm,bai2023accelerating,yang2023diffsound,vall-e,musicgen,wang2023speechx,BASETTS}. When it comes to streaming applications, autoregressive language model-based approaches have an advantage as they can generate audio with low latency by processing data sequentially, while diffusion models have to rely on successive non-autoregressive queries to reconstruct audio.

To leverage the language model's capability in the audio domain, vector quantization has been used to represent audio signals as discrete codes. While some works \cite{dang2022training,BASETTS,kharitonov2023speak} 
leverage tokens obtained via self-supervised pretraining, which can be fused with speaker information during generation, other works \cite{vall-e,musicgen} rely on tokens from an audio codec trained with residual vector quantization (RVQ) \cite{soundstream,encodec,dac}, which can be solely used to synthesize the raw audio. AudioLM \cite{audiolm} proposes using autoregressive transformers to generate one token per step, where coarse and fine acoustic tokens are modeled separately. VALL-E \cite{vall-e} models the first token in each audio frame with an autoregressive transformer, and sequentially predicts the second to the last codes for all frames using a shared non-autoregressive transformer. MusicGen \cite{musicgen} proposes a delayed generation pattern that can parallelize 4-code generation in each step. 
In this work, we adopt the delayed generation pattern in MusicGen, with proposed techniques to enable high-fidelity and low-latency speech generation.

\section{Background}
\label{sec:background}

In this section, we go over some key concepts in audio tokenization and how to use them in audio generation.

\paragraph*{Audio Compression with Residual Vector Quantization}
\vspace{-10pt}
A pivotal element for applying language models in the audio domain is an audio tokenization component, which usually comprises an encoder, a quantizer, and a decoder \cite{soundstream,encodec,dac,tfcodec}. The encoder transforms the audio into a latent speech representation of $T$ time steps $\bz_1, \bz_2, ..., \bz_T$, which is recursively quantized by a sequence of quantizers to produce $Q$ codes $\bc_i=[c_{i}^{(1)}, c_{i}^{(2)}, ..., c_{i}^{ (Q)}]\in\mathcal{C}^Q$ for each $\bz_i$, where $\mathcal{C}$ is the set of codebook indices. %
As a result, the first few codes mainly represent the content of the audio, while the later codes represent fine-grained details \cite{vall-e}.%
We refer to codes in early codebooks as \textit{high-level}, and codes in later codebooks as \textit{low-level}.

\vspace{-10pt}
\paragraph*{Audio Generation via Discrete Tokens}

The compression rate and reconstruction quality of RVQ codes inspire a number of works to formulate audio generation as a language modeling task; however predicting codes sequentially poses a challenge of high inference time. MusicGen \cite{musicgen} proposes reducing the context size by predicting $Q$ codes together, which is made possible by shifting the codebooks so that each step predicts $Q$ codes, but only one comes from each frame. Let $\boldsymbol{C'}=\left[\bc'_1,...,\bc'_{T'}\right]$ be the shifted codes obtained from $\boldsymbol{C}$, where $T'=T+Q-1$ is the length of the shifted code sequence. We have $\bc'_i=\left[c_{i}^{(1)}, \dots, c_{i-(Q-1)}^{(Q)}\right]$, assuming padding values for invalid code positions. $\boldsymbol{C'}$ is modeled by the transformer instead of $\boldsymbol{C}$. As a result, the $i$-th audio frame in $\boldsymbol{C}$ is fully generated at the $(i+Q-1)$-th step in $\boldsymbol{C'}$. This provides an inexact autoregressive decomposition to model the distribution of discrete codes. Although the performance of the delayed pattern is still behind that of the flatten pattern, it produces reasonable balance between the audio quality and the inference budget.
\vspace{-10pt}
\paragraph*{The Content Accuracy vs Voice Quality Trade-off}

With the delayed generation pattern, the model needs to distribute its capacity across all codebooks to produce one code from each of them. With limited model capacity, prioritizing some codebooks may lead to the poor prediction of other codebooks, which results in a content accuracy - voice quality trade-off. In Table \ref{tab:tradeoff}, we show that training two models that shares the architecture, one being assigned equal weights to all codebook losses and the other being assigned higher weights for the high-level codebooks, does not yield satisfactory scores for both aspects in either of these models. In the following section, we propose several approaches to mitigate this issue.%

\begin{table}[h]
\caption{Content accuracy (represented by CER, WER) - voice quality (represented by SS, O-MOS) trade-off when adjusting the codebook priority. Metric details are provided in Section \ref{sec:experiments}.}
\vskip 0.15in
\begin{center}
\begin{small}
\setlength{\tabcolsep}{4pt}
    \centering
    \begin{tabular}{lcccccccc}
        \toprule
        Focus on & CER ($\downarrow$) & WER ($\downarrow$) & SS ($\uparrow$) & O-MOS ($\uparrow$) \\ \midrule
        None & 12.4 & 23.8 &  \textbf{58.5} & \textbf{3.66} \\
        High-level codes & \textbf{2.4} & \textbf{4.7} & 49.4 & 3.33 \\ \bottomrule
    \end{tabular}
    \label{tab:tradeoff}
\end{small}
\end{center}
\vskip -0.1in
\end{table}

\section{Our Proposed Model}
\label{sec:proposed}

In this section, we describe our model and two proposed techniques to efficiently predict all codebooks in a decoding step. For convenience, we use $[c_1,\dots,c_T]$ instead of $[c'_1,\dots,c'_{T'}]$ to denote input tokens for each transformer step during training.

\begin{figure*}
    \centering
    \includegraphics[width=\textwidth]{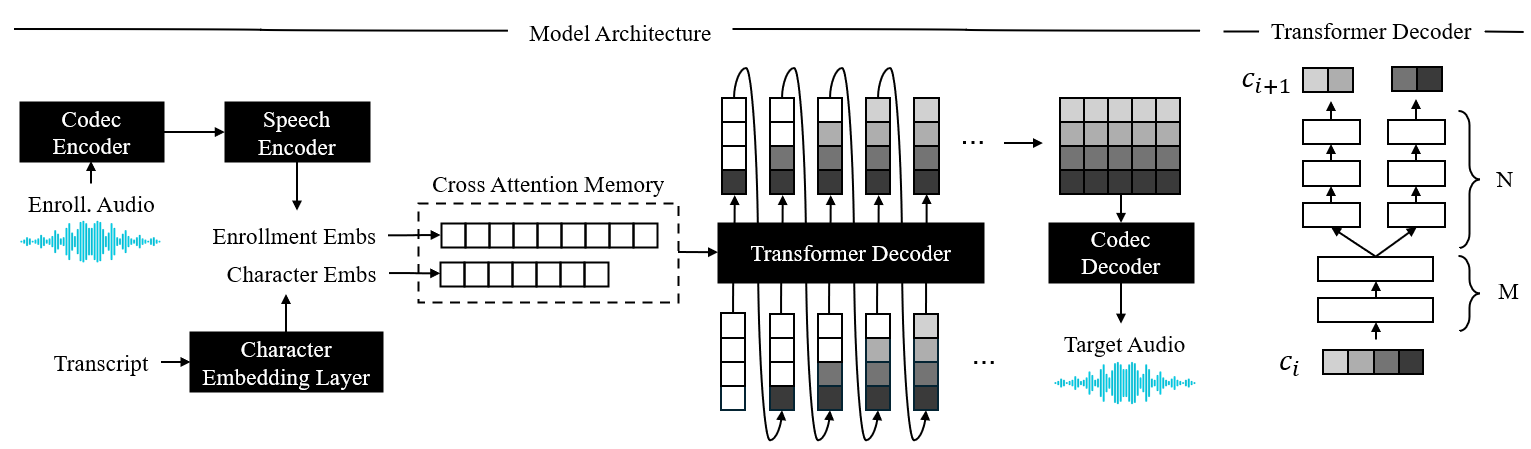}
    \caption{(Left) Our proposed architecture. 
    Our model consists of a neural audio codec to convert between waveforms and discrete codes, a speech encoder to infer enrollment embeddings, and a transformer decoder to generate discrete tokens from conditions.  
    (Right) Transformer decoder with parallel codebook group heads}
    \label{fig:architecture}
\end{figure*}

\subsection{Model Architecture}

Our model shares the architecture with a GPT-style autoregressive language model. It consists of a \textit{neural audio codec} that encodes raw audio to codes and decodes codes back to raw audio, a \textit{speech encoder} and a \textit{text embedding layer} to provides voice and text condition vectors, respectively, and a \textit{transformer decoder} to generate audio tokens. For the speech encoder, we employ an encoder-decoder transformer, which takes a variable-length enrollment speech and produces a fixed-length sequence of features in a non-autoregressive manner. %
The main transformer decoder processes $Q$ codes from $Q$ codebooks in each time step. The decoder takes a sum of all code embeddings $\bx_t=\sum_{q=1}^Q \text{Emb}_q\left(c_t^{(q)}\right)$ and predicts all codes from a vector output $\bp_t^{(q)}=\text{Softmax}(\text{Proj}_q\left(\bo_t\right))$, where $\text{Emb}_q, \text{Proj}_q$ are the embedding and projection layer for the $q$-th codebook, $\bx_t$, $\bo_t$ are the input and output for the transformer step $t$, $\bp_t^{(q)}$ is the softmax probability distribution of the $q$-th code at the step $t$. The input and target codes are shifted for delayed generation, similar to MusicGen \cite{musicgen}. Figure \ref{fig:architecture} (left) illustrates the end-to-end architecture of our model.

\subsection{Adaptive Codebook Weights}
\label{sec:adaptive}
To address the content accuracy - voice quality tradeoff, we propose an adaptive codebook weighing technique that enables the model to redistribute its capacity for each codebook during training. Since high-level codes contribute more to the final constructed frame and guide the content of the speech, we want to prioritize them at the early training stage. As the accuracy for high-level codebooks improves, we can focus more on lower-level codes that are harder to predict correctly. We propose a mechanism to fine-tune the model's focus down to the frame level: we assign a weight for each term in the loss based on how well higher-level codes in the same frame are predicted. Let $\tilde{p}_t^{(q)}=\bp_{t}^{(q)}\left[c_{t}^{(q)}\right]$ be the softmax probability value for correctly predicting the code $c_{t}^{(q)}$. The weighted loss is defined as \begin{equation}
\mathcal{L}=\frac{1}{TQ}\sum_{q=1}^{Q}\sum_{t=1}^T\overline{w}_t^{(q)}\mathcal{L}_{CE}\left(\bp_{t}^{(q)}, c_{t}^{(q)}\right),
\end{equation}where $\overline{w}_{t}^{(1)}=1$, $\overline{w}_{t}^{(q>1)}=\prod_{q'<q}\left(\Tilde{p}_{t}^{(q')}\right)^{\lambda}$ being the weight associated with the loss of predicting $c_t^{(q)}$, $\lambda\ge 0$ is a hyperparameter controlling the decay rate as we get to lower level codebooks. The bar in $\overline{w}$ indicates that it does not allow the gradients to backpropagate through. When $\lambda=0$, all codebook loss terms are given the same weight. When $\lambda>0$, the codebook weights at each time step $t$ strictly decrease; the decreasing rate depends on the probability of predicting correctly for all previous codes in the same frame. This is applied recursively throughout all the codes predicted in each audio frame, and applied differently to each audio frame in the target speech. In general, this loss encourages the model to focus on high-level codes at the beginning and shifts the focus to lower-level codes as training progresses.
To mitigate the weight vanishing for low-level codes, we also introduce a threshold $p_{\max}$ and ignore the $q$-th code if the probability of correctly predicting the code $\tilde{\bp}_t^{(q)}$ is greater than $p_{\max}$. Weights for the remaining codes in the same frame are scaled such that the largest weight becomes 1. This allows training to ignore easy predictions.

\subsection{Parallel Codebook Group Heads}

Since the transformer needs to predict codes in all codebooks within a single step, we propose enhancing the modeling capacity of each step by grouping $Q$ codes into $G$ groups and predict codes in each group together in parallel decoding steps with their own hidden representations. Besides relaxing the number of codes needed to be predicted from a query, grouping codes also allows each group to attend to different parts in the memory, for example, low-level codes may benefit more from codes generated in recent time steps. Let $L$ be the number of transformer layers, we keep $M$ shared layers for all groups of codebooks and use $N=L-M$ layers to process each group independently. At the transition layer, we split the layer output $\bo_{t,M}$ into $G$ next layer inputs by using group-specific projection layers: $\bh^g_{t,M+1}=\text{GProj}_g(\bo_{t,M})$, where $g$ is the group index. At the last layer, we obtain the probability for each code $c_t^{(q)}$ from the output of the corresponding group $\gamma(q)$ to the $q$-th codebook as $\bp_t^{(q)}=\text{Softmax}\left(\text{Proj}_q\left(\bo^{\gamma(q)}_{t,L}\right)\right)$. Figure \ref{fig:architecture} (Right) gives an example of the transformer decoder when $M=2$, $N=3$, $Q=4$, and $G=2$. These group specific layers only slightly increase the model size, although the inference time and memory increase similarly to when increasing the batch size by $G$ times for the last $N$ layers. On capable hardware, this may have an insignificant impact on the speed due to parallelization.

\section{Experiments}

\label{sec:experiments}
\subsection{Setup}
\paragraph*{Model Architecture} For audio codec, we use Encodec 24kHz \cite{encodec} at 12kbps compression rate and 75fps where each frame is represented by 16 codes of total 160 bits. Our transformer decoder consists of 12 layers, each having 16 heads, with a layer dimension of 1,536 and a feedforward dimension of 6,144. The speech encoder has a non-autoregressive transformer architecture of 6 layers, 8 heads, and a hidden and output dimension of 1,024, which takes continuous speech features from the Encodec and outputs a sequence of 64 vector features. For models with parallel codebook group heads, we group 16 codes into 8 groups of two each. The decoder has the same configuration of 12 transformer layers with the first $M=6$ layers processing frame features and the last $N=6$ layers processing group features in parallel. The size of the model without/with group heads is 581M/615M, including 77M parameters for the speech encoder that is not used during decoding.

\vspace{-10pt}
\paragraph*{Dataset} We pretrain our model on LibriLight \cite{librilight}, a large unlabelled speech corpus of 60k hours. Since transcripts are not available, we employ ASR models to derive the transcript of all audio segments in the dataset. Specifically, we split recordings by each speaker into segments of 160 to 200 seconds, and use Wav2Vec 2.0 Large (LV-60) + Self Training \cite{wav2vec2} to extract character-base transcripts.
Each audio segment in the batch has a duration ranging from 0.1 to 10 seconds, which is sampled such that it start and end with a complete word based on its time-align grapheme sequence. For better data loading efficiency during training, audio is only stored and available in the form of codec codes extracted by Encodec \cite{encodec}. We use the dev-clean set of LibriTTS \cite{LibriTTS} as the validation set and select checkpoints based on the value of CER on the validation set. Our test set is derived from the test-clean set of LibriTTS, where we only keep audio of 1-10s duration and randomly sample an enrollment speech of 3-5s audio within 50s from the target audio. This results in a test set of 4.4 hours of 3,624 samples.
\begin{table*}[t]
\caption{Comparison between our models and baselines when using 3s and 5s of enrollment audio. For reference, we also include results from industrial baselines with access to more data and may be optimized for the inference speed.}

\vskip 0.15in
\begin{center}
\begin{small}
\setlength{\tabcolsep}{4.5pt}
    \centering
    \begin{tabular}{lccccccccccccc}
        \toprule
          & \multicolumn{2}{c}{CER ($\downarrow$)} & 
            \multicolumn{2}{c}{WER ($\downarrow$)} & \multicolumn{2}{c}{PER ($\downarrow$)} & 
            \multicolumn{2}{c}{SS ($\uparrow$)} & \multicolumn{2}{c}{O-MOS ($\uparrow$)} & S-MOS ($\uparrow$) & RTF ($\downarrow$) & Lat. ($\downarrow$) \\ 
            & 3s & 5s & 3s & 5s & 3s & 5s & 3s & 5s & 3s & 5s & 5s & 5s & 5s\\
        \midrule
        Reference &  \multicolumn{2}{c}{1.2} & 
            \multicolumn{2}{c}{2.7} & \multicolumn{2}{c}{12.3} & 
            \multicolumn{2}{c}{76.7} & \multicolumn{2}{c}{3.80} & - & - & - \\
        Reference (16-code, 12 kbps) &
        \multicolumn{2}{c}{1.4} &
        \multicolumn{2}{c}{2.9} & \multicolumn{2}{c}{12.4} & 
        \multicolumn{2}{c}{71.1} & \multicolumn{2}{c}{3.72} & 0.00 & - & - \\
        Reference (8-code, 6 kbps) & \multicolumn{2}{c}{1.6} & 
            \multicolumn{2}{c}{2.9} & \multicolumn{2}{c}{12.4} &
            \multicolumn{2}{c}{67.8} & \multicolumn{2}{c}{3.63} & -0.03 & - & - \\
        \midrule
        \textit{Industrial Baselines} \\
        XTTS-v1 & 2.2 & 2.0 & 6.3 & 5.9 & 12.3 & 12.2 & 48.2 & 50.5 & 3.93 & 3.93 & - & - & - \\
        XTTS-v2 & 2.1 & 2.0 & 
            7.0 & 6.5 & 12.7 & 12.4 & 
            57.0 & 60.3 & 3.81 & 3.83 & - & 0.43 & 0.36 \\
        MetaVoice-1B & 7.9 & 6.7 & 14.0 & 12.7 & 19.4 & 18.8 & 53.9 & 56.6 & 3.60 & 3.60 & - & 2.33 & - \\
        \midrule
        \textit{Baselines} \\
        YourTTS \cite{yourtts} & 4.8 & 4.6 & 8.9 & 8.8 & 15.3 & 15.4 & 46.4 & 48.5 & \textbf{3.71} & \textbf{3.72} & -0.26 & \textbf{0.06} & - \\
        VALL-E (SpeechX ft) \cite{vall-e,wang2023speechx} & 4.0 & 3.9 & 
            6.4 & 6.0 & 15.2 & 14.9 & 53.0 & 58.0 & 3.69 & 3.70 & -0.12  &
            0.87 & - \\

        \midrule
        Ours - $\lambda=0$ & 12.4 & 12.4 & 24.0 & 23.8 & 23.2 &  23.1 & 55.5 & 58.5 & 3.63 & 3.66 & -0.19 & 0.87 & \textbf{0.19} \\
        Ours - $\lambda=0.1$ & 3.5 & 3.6 & 6.8 & 7.0 & 13.9 & 14.0 & 54.4 & 57.1 & 3.57 & 3.57 & -0.34 & 0.87 & \textbf{0.19} \\
        \hspace{5pt}+ $p_{\max}=0.5$  & \textbf{3.0} & \textbf{3.0} & \textbf{6.1} & \textbf{6.0} & \textbf{13.4} & \textbf{13.3} & 55.1 & 57.6 & 3.59 & 3.59 & -0.14 & 0.87 & \textbf{0.19} \\
        Ours - 8 groups, $\lambda=0.05$ & 3.7 & 3.7 & 7.2 & 6.9 & 14.4 & 14.1 & \textbf{56.8} & \textbf{59.5} & 3.66 & 3.66 & -0.04 & 0.96 & 0.20 \\ %
        \hspace{5pt}+ DeepFilterNet \cite{DeepFilterNet} & 3.8 & 3.5 & 7.2 & 6.8 & 14.2 & 14.1 & 56.3 & 58.9 & 3.71 & 3.71 & \textbf{+0.01} & 0.97 & 0.21 \\
        \bottomrule
    \end{tabular}
    \label{tab:main_res}

\end{small}
\end{center}
\vskip -0.1in
\end{table*}
\vspace{-10pt}
\paragraph*{Training \& Inference}

Our models are trained with a batch size of 64 for 1M steps or batch size of 16 for around 3M steps for models with codebook group heads on 4 A100 GPUs.
We select the checkpoint based on a validation set of 500 samples taken from the dev-clean set of LibriTTS. We do a beam search to scan for the decoding temperature $\tau_{sb}\in[1.0, 1.1, 1.2]$, the number of sample-based codes $n_{sb}=[1, 2, 3, 4, 8, 16]$, the top-$k$ sampling's parameter $k\in[10,15,20]$ for \textit{each} codebook, and choose the best hyperparameters based on the value of $(\text{SS}-\text{CER})$ on a validation set of 40 samples. For adaptive codebook weights, we report results with $\lambda=0.1$, with and without the probability threshold $p_{\max}=0.5$. For models with parallel codebook group heads, we report results with $\lambda=0.05$. We also include results when using an enhancer \cite{DeepFilterNet} on the generated speech.

\vspace{-10pt}
\paragraph*{Objective Evaluation}
\label{sec:metrics}
We evaluate output audios in terms of (1) transcript error rates (TER), (2) speaker similarity scores (SS), and (3) objective perceptual speech quality score P.808 (O-MOS) \cite{reddy2022dnsmos}. For (1), we report character error rate (CER)\footnote{\scriptsize\url{hf.co/facebook/wav2vec2-base-960h} \cite{wav2vec2}}, phoneme error rate (PER)\footnote{\scriptsize{\url{hf.co/facebook/wav2vec2-xlsr-53-espeak-cv-ft}} \cite{wav2vec2-phn}}, and word error rate (WER)\footnote{\scriptsize{\url{hf.co/hubert-large-ls960-ft}} \cite{hubert}}. 
For (2), we report speaker similarity scores by computing the cosine similarity between speaker embeddings obtained from ECAPA-TDNN model\footnote{\scriptsize{\url{hf.co/speechbrain/spkrec-ecapa-voxceleb}} \cite{ecapa,speechbrain}}.
We only compute scores over samples where the reference audio is longer than 3s, and against the full-length utterance where the enrollment speech is extracted from.
All clips are sampled to 16kHz for evaluation. We also simulate real-time inference and measure the speed in terms of the real-time factor (RTF) and the latency (Lat) on 1 NVIDIA RTX 6000 Ada Generation GPU. Our latency excludes the time for the computation of the speaker condition, which can be cached for the same speaker. We do not report the latency for non-streaming models, in which cases the latency depends on the generated audio duration.
\vspace{-10pt}

\paragraph*{Subjective Evaluation} We conduct subjective evaluation and report the relative Mean Opinion Score (S-MOS) on uniformly sampled 148 utterances (around 11 mins in total) from the test set. Each subject is asked to rate the quality of the audio on a scale of 1-5. Each audio is evaluated by 7 subjects.
\vspace{-10pt}

\paragraph*{Baselines}

We compare our models with YourTTS \cite{yourtts} (87M) and VALL-E \cite{vall-e} (488M) using pretrained checkpoints. The VALL-E checkpoint is taken from the SpeechX \cite{wang2023speechx}, which is reported with better performance than the original VALL-E through multitask finetuning. We also include the results of industrial baselines such as XTTS-v2\footnote{\scriptsize\url{huggingface.co/coqui/XTTS-v2}} and MetaVoice-1B\footnote{\scriptsize\url{huggingface.co/metavoiceio/metavoice-1B-v0.1}}, whose training details and datasets are not published. %

\subsection{Results}

\paragraph*{Speech Quality} Table \ref{tab:main_res} compares our models to other baselines. Our TTS adapted MusicGen model ($\lambda=0$) performs well on the SS metric; however CER/WER/PER (or TER) scores are noticeably high. A large improvement in TER scores is achieved when using adaptive codebook weights; however, SS and MOS scores are also affected. Our model with $p_{\max}=0.5$ or 8 codebook groups further improves these scores to be better than or comparable to baselines. In terms of subjective scores, our 8-group model shows better quality than the baseline systems and comes close to the 6kbps compressed reference audio, with the special case when using an enhancer where no drop is observed compared to the 12kbps reference (upper bound). Samples are available at \url{trungd.github.io/livespeech}.

\vspace{-10pt}

\paragraph*{Speed \& Latency} Our RTF is comparable to VALL-E's, despite being fully auto-regressive. %
The model with 8 groups has RTF increased only by 0.09s or 10\%, showing the efficiency of parallelization. Our model operates with a delay of 200ms, making it suitable for low-latency applications.

\section{Conclusion}

We present LiveSpeech, a fully autoregressive zero-shot text-to-speech model that enables live streaming of output audio. The proposed techniques, including adaptive codebook loss weights and parallel processing of codebook groups, show competitive performance and successfully address the challenges of existing systems in a real-time or low-latency setting. %

\newpage
\bibliographystyle{IEEEtran}
\bibliography{strings}

\begin{thebibliography}{10}
\providecommand{\url}[1]{#1}
\csname url@samestyle\endcsname
\providecommand{\newblock}{\relax}
\providecommand{\bibinfo}[2]{#2}
\providecommand{\BIBentrySTDinterwordspacing}{\spaceskip=0pt\relax}
\providecommand{\BIBentryALTinterwordstretchfactor}{4}
\providecommand{\BIBentryALTinterwordspacing}{\spaceskip=\fontdimen2\font plus
\BIBentryALTinterwordstretchfactor\fontdimen3\font minus \fontdimen4\font\relax}
\providecommand{\BIBforeignlanguage}[2]{{%
\expandafter\ifx\csname l@#1\endcsname\relax
\typeout{** WARNING: IEEEtran.bst: No hyphenation pattern has been}%
\typeout{** loaded for the language `#1'. Using the pattern for}%
\typeout{** the default language instead.}%
\else
\language=\csname l@#1\endcsname
\fi
#2}}
\providecommand{\BIBdecl}{\relax}
\BIBdecl

\bibitem{vall-e}
C.~Wang, S.~Chen, Y.~Wu, Z.~Zhang, L.~Zhou, S.~Liu, Z.~Chen, Y.~Liu, H.~Wang, J.~Li \emph{et~al.}, ``Neural codec language models are zero-shot text to speech synthesizers,'' \emph{arXiv preprint arXiv:2301.02111}, 2023.

\bibitem{vall-ex}
Z.~Zhang, L.~Zhou, C.~Wang, S.~Chen, Y.~Wu, S.~Liu, Z.~Chen, Y.~Liu, H.~Wang, J.~Li \emph{et~al.}, ``Speak foreign languages with your own voice: Cross-lingual neural codec language modeling,'' \emph{arXiv preprint arXiv:2303.03926}, 2023.

\bibitem{musicgen}
J.~Copet, F.~Kreuk, I.~Gat, T.~Remez, D.~Kant, G.~Synnaeve, Y.~Adi, and A.~D{\'e}fossez, ``Simple and controllable music generation,'' \emph{arXiv preprint arXiv:2306.05284}, 2023.

\bibitem{naturalspeech2}
K.~Shen, Z.~Ju, X.~Tan, Y.~Liu, Y.~Leng, L.~He, T.~Qin, S.~Zhao, and J.~Bian, ``Naturalspeech 2: Latent diffusion models are natural and zero-shot speech and singing synthesizers,'' \emph{arXiv preprint arXiv:2304.09116}, 2023.

\bibitem{audiolm}
Z.~Borsos, R.~Marinier, D.~Vincent, E.~Kharitonov, O.~Pietquin, M.~Sharifi, D.~Roblek, O.~Teboul, D.~Grangier, M.~Tagliasacchi \emph{et~al.}, ``Audiolm: a language modeling approach to audio generation,'' \emph{IEEE/ACM Transactions on Audio, Speech, and Language Processing}, 2023.

\bibitem{soundstream}
N.~Zeghidour, A.~Luebs, A.~Omran, J.~Skoglund, and M.~Tagliasacchi, ``Soundstream: An end-to-end neural audio codec,'' \emph{IEEE/ACM Transactions on Audio, Speech, and Language Processing}, vol.~30, pp. 495--507, 2021.

\bibitem{tacotron2}
J.~Shen, R.~Pang, R.~J. Weiss, M.~Schuster, N.~Jaitly, Z.~Yang, Z.~Chen, Y.~Zhang, Y.~Wang, R.~Skerrv-Ryan \emph{et~al.}, ``Natural tts synthesis by conditioning wavenet on mel spectrogram predictions,'' in \emph{2018 IEEE international conference on acoustics, speech and signal processing (ICASSP)}.\hskip 1em plus 0.5em minus 0.4em\relax IEEE, 2018, pp. 4779--4783.

\bibitem{ping2017deep}
W.~Ping, K.~Peng, A.~Gibiansky, S.~O. Arik, A.~Kannan, S.~Narang, J.~Raiman, and J.~Miller, ``Deep voice 3: Scaling text-to-speech with convolutional sequence learning,'' \emph{arXiv preprint arXiv:1710.07654}, 2017.

\bibitem{wu2022adaspeech}
Y.~Wu, X.~Tan, B.~Li, L.~He, S.~Zhao, R.~Song, T.~Qin, and T.-Y. Liu, ``Adaspeech 4: Adaptive text to speech in zero-shot scenarios,'' \emph{arXiv preprint arXiv:2204.00436}, 2022.

\bibitem{casanova2021sc}
E.~Casanova, C.~Shulby, E.~G{\"o}lge, N.~M. M{\"u}ller, F.~S. De~Oliveira, A.~C. Junior, A.~d.~S. Soares, S.~M. Aluisio, and M.~A. Ponti, ``Sc-glowtts: An efficient zero-shot multi-speaker text-to-speech model,'' pp. 3645--3649, 2021.

\bibitem{mqtts}
L.-W. Chen, S.~Watanabe, and A.~Rudnicky, ``A vector quantized approach for text to speech synthesis on real-world spontaneous speech,'' \emph{arXiv preprint arXiv:2302.04215}, 2023.

\bibitem{bai2023accelerating}
Y.~Bai, T.~Dang, D.~Tran, K.~Koishida, and S.~Sojoudi, ``Accelerating diffusion-based text-to-audio generation with consistency distillation,'' \emph{arXiv preprint arXiv:2309.10740}, 2023.

\bibitem{yang2023diffsound}
D.~Yang, J.~Yu, H.~Wang, W.~Wang, C.~Weng, Y.~Zou, and D.~Yu, ``Diffsound: Discrete diffusion model for text-to-sound generation,'' \emph{IEEE/ACM Transactions on Audio, Speech, and Language Processing}, 2023.

\bibitem{wang2023speechx}
X.~Wang, M.~Thakker, Z.~Chen, N.~Kanda, S.~E. Eskimez, S.~Chen, M.~Tang, S.~Liu, J.~Li, and T.~Yoshioka, ``Speechx: Neural codec language model as a versatile speech transformer,'' \emph{arXiv preprint arXiv:2308.06873}, 2023.

\bibitem{BASETTS}
M.~Lajszczak, G.~C. Ruiz, Y.~Li, F.~Beyhan, A.~van Korlaar, F.~Yang, A.~Joly, Álvaro Martín~Cortinas, A.~Abbas, A.~Michalski, A.~Moinet, S.~Karlapati, E.~Muszynska, H.~Guo, B.~Putrycz, S.~L. Gambino, K.~Yoo, E.~Sokolova, and T.~Drugman, ``Base tts: Lessons from building a billion-parameter text-to-speech model on 100k hours of data,'' \emph{arXiv}, 2024.

\bibitem{dang2022training}
T.~Dang, D.~Tran, P.~Chin, and K.~Koishida, ``Training robust zero-shot voice conversion models with self-supervised features,'' in \emph{ICASSP 2022-2022 IEEE International Conference on Acoustics, Speech and Signal Processing (ICASSP)}.\hskip 1em plus 0.5em minus 0.4em\relax IEEE, 2022, pp. 6557--6561.

\bibitem{kharitonov2023speak}
E.~Kharitonov, D.~Vincent, Z.~Borsos, R.~Marinier, S.~Girgin, O.~Pietquin, M.~Sharifi, M.~Tagliasacchi, and N.~Zeghidour, ``Speak, read and prompt: High-fidelity text-to-speech with minimal supervision,'' \emph{Transactions of the Association for Computational Linguistics}, vol.~11, pp. 1703--1718, 2023.

\bibitem{encodec}
A.~D{\'e}fossez, J.~Copet, G.~Synnaeve, and Y.~Adi, ``High fidelity neural audio compression,'' \emph{arXiv preprint arXiv:2210.13438}, 2022.

\bibitem{dac}
R.~Kumar, P.~Seetharaman, A.~Luebs, I.~Kumar, and K.~Kumar, ``High-fidelity audio compression with improved rvqgan,'' \emph{arXiv preprint arXiv:2306.06546}, 2023.

\bibitem{tfcodec}
X.~Jiang, X.~Peng, Y.~Zhang, and Y.~Lu, ``Disentangled feature learning for real-time neural speech coding,'' in \emph{ICASSP 2023-2023 IEEE International Conference on Acoustics, Speech and Signal Processing (ICASSP)}.\hskip 1em plus 0.5em minus 0.4em\relax IEEE, 2023, pp. 1--5.

\bibitem{librilight}
J.~Kahn, M.~Rivi{\`e}re, W.~Zheng, E.~Kharitonov, Q.~Xu, P.-E. Mazar{\'e}, J.~Karadayi, V.~Liptchinsky, R.~Collobert, C.~Fuegen \emph{et~al.}, ``Libri-light: A benchmark for asr with limited or no supervision,'' in \emph{ICASSP 2020-2020 IEEE International Conference on Acoustics, Speech and Signal Processing (ICASSP)}.\hskip 1em plus 0.5em minus 0.4em\relax IEEE, 2020, pp. 7669--7673.

\bibitem{wav2vec2}
A.~Baevski, Y.~Zhou, A.~Mohamed, and M.~Auli, ``wav2vec 2.0: A framework for self-supervised learning of speech representations,'' \emph{Advances in neural information processing systems}, vol.~33, pp. 12\,449--12\,460, 2020.

\bibitem{LibriTTS}
H.~Zen, V.~Dang, R.~A.~J. Clark, Y.~Zhang, R.~J. Weiss, Y.~Jia, Z.~Chen, and Y.~Wu, ``Libritts: A corpus derived from librispeech for text-to-speech,'' in \emph{Interspeech}, 2019.

\bibitem{yourtts}
E.~Casanova, J.~Weber, C.~D. Shulby, A.~C. Junior, E.~G{\"o}lge, and M.~A. Ponti, ``Yourtts: Towards zero-shot multi-speaker tts and zero-shot voice conversion for everyone,'' in \emph{International Conference on Machine Learning}.\hskip 1em plus 0.5em minus 0.4em\relax PMLR, 2022, pp. 2709--2720.

\bibitem{DeepFilterNet}
H.~Schröter, A.~N. Escalante-B., T.~Rosenkranz, and A.~Maier, ``{DeepFilterNet}: A low complexity speech enhancement framework for full-band audio based on deep filtering,'' in \emph{ICASSP 2022 IEEE International Conference on Acoustics, Speech and Signal Processing (ICASSP)}.\hskip 1em plus 0.5em minus 0.4em\relax IEEE, 2022.

\bibitem{reddy2022dnsmos}
C.~K. Reddy, V.~Gopal, and R.~Cutler, ``Dnsmos p. 835: A non-intrusive perceptual objective speech quality metric to evaluate noise suppressors,'' in \emph{ICASSP 2022-2022 IEEE International Conference on Acoustics, Speech and Signal Processing (ICASSP)}.\hskip 1em plus 0.5em minus 0.4em\relax IEEE, 2022, pp. 886--890.

\bibitem{wav2vec2-phn}
Q.~Xu, A.~Baevski, and M.~Auli, ``Simple and effective zero-shot cross-lingual phoneme recognition,'' \emph{arXiv preprint arXiv:2109.11680}, 2021.

\bibitem{hubert}
W.-N. Hsu, B.~Bolte, Y.-H.~H. Tsai, K.~Lakhotia, R.~Salakhutdinov, and A.~Mohamed, ``Hubert: Self-supervised speech representation learning by masked prediction of hidden units,'' \emph{IEEE/ACM Transactions on Audio, Speech, and Language Processing}, vol.~29, pp. 3451--3460, 2021.

\bibitem{ecapa}
B.~Desplanques, J.~Thienpondt, and K.~Demuynck, ``{ECAPA-TDNN:} emphasized channel attention, propagation and aggregation in {TDNN} based speaker verification,'' in \emph{Interspeech 2020}.\hskip 1em plus 0.5em minus 0.4em\relax {ISCA}, 2020, pp. 3830--3834.

\bibitem{speechbrain}
M.~Ravanelli, T.~Parcollet, P.~Plantinga, A.~Rouhe, S.~Cornell, L.~Lugosch, C.~Subakan, N.~Dawalatabad, A.~Heba, J.~Zhong, J.-C. Chou, S.-L. Yeh, S.-W. Fu, C.-F. Liao, E.~Rastorgueva, F.~Grondin, W.~Aris, H.~Na, Y.~Gao, R.~D. Mori, and Y.~Bengio, ``{SpeechBrain}: A general-purpose speech toolkit,'' 2021, arXiv:2106.04624.

\bibitem{MBD}
R.~S. Roman, Y.~Adi, A.~Deleforge, R.~Serizel, G.~Synnaeve, and A.~Defossez, ``From discrete tokens to high-fidelity audio using multi-band diffusion,'' \emph{Advances in neural information processing systems}, 2023.

\bibitem{reddy2021dnsmos}
C.~K. Reddy, V.~Gopal, and R.~Cutler, ``Dnsmos: A non-intrusive perceptual objective speech quality metric to evaluate noise suppressors,'' in \emph{ICASSP 2021 IEEE International Conference on Acoustics, Speech and Signal Processing (ICASSP)}.\hskip 1em plus 0.5em minus 0.4em\relax IEEE, 2021, pp. 6493--6497.

\end{thebibliography}

\clearpage
\appendix

\section{Appendix}

\subsection{Background}
\label{sec:appendix:background}
\subsubsection{Audio Compression with Residual Vector Quantization}

A pivotal element for applying language models in the audio domain is an audio tokenization component. An encoder (usually a multi-layer convolutional encoder) encodes the audio into a latent speech representation of $T$ time steps $\bz_1, \bz_2, ..., \bz_T\in\mathcal{Z}$.
A residual vector quantizer is a sequence of $Q$ quantizers that recursively quantize the residual from feature vectors $\bz_i$ to codes 
$\bc_i=[c_{i}^{(1)}, c_{i}^{(2)}, ..., c_{i}^{ (Q)}]\in\mathcal{C}^Q$
, where $\mathcal{C}$ is the set of codebook indices. For some feature $\bz$, this is processed as follows: $\br_1=\bz, c^{(q)}=\arg\min_k(||\br^{(q)}-\bd_{k}^{(q)}||), \br^{(q+1)}=\br^{(q)}-\bd^{(q)}_{c^{(q)}}$ for $q\in[1, Q]$, where $\br^{(q)}$ is the residual to quantize at the $q$-th quantizer. When applying this to all time steps $\bz_1, \bz_2, ..., \bz_T$, the original signal is encoded as $\boldsymbol{C}\in\mathcal{C}^{Q\times T}$.

When the discrete audio representation is learned through a self-reconstruction task, it becomes a neural audio codec that can be used to compress an audio to a very low bit-rate \cite{soundstream,encodec}. The training is usually aided with neural network discriminators to improve the reconstruction quality.

\begin{figure}[ht]
    \centering
    \includegraphics[width=0.49\textwidth]{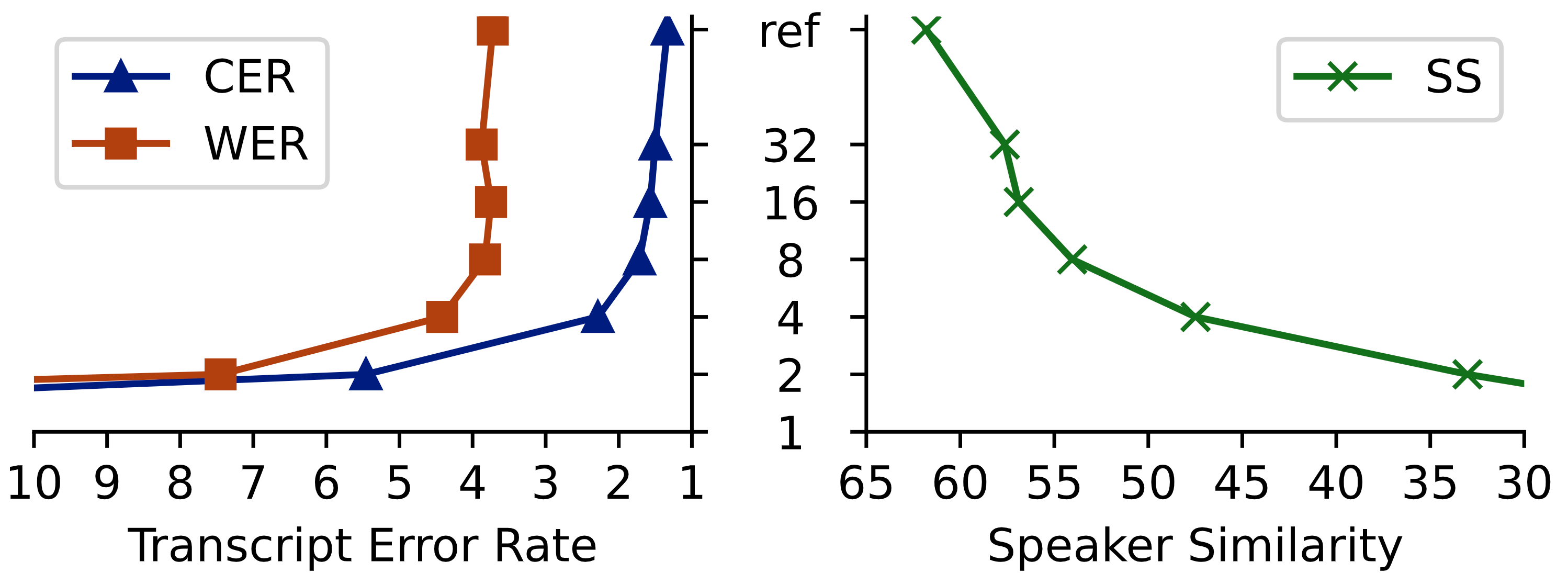}
    \caption{Transcript error rates (CER, WER) and speaker similarity scores (SS$_{e}$) of the reference audio decoded by the number of codebooks used. `ref' represents the original audio. Metrics are given details in Section \ref{sec:metrics}}
    \label{fig:ref_by_num_codes}
\end{figure}

This provides audio codes that can be generated autoregressively, since in a single frame, each code only depends on previous codes. As a result, the first few codes mainly represent the content of the audio, while the later codes represent fine-grained details. This is shown in Figure \ref{fig:ref_by_num_codes}, where the audio reconstructed from the first two codes are already able to achieve around 5\% CER, while the speaker similarity continues to benefit from an increasing number of codebooks. We refer to codes in early codebooks as \textit{high-level} codes, and codes in later codebooks as \textit{low-level} codes.

\subsubsection{Audio Generation via Discrete Tokens}

\begin{table}[t]
\caption{Details of audio tokenizers Encodec \cite{encodec}, TF-Codec \cite{tfcodec}, DAC \cite{dac} with their compression bit rate (BR, kbps) and their scores on zero-shot TTS metrics. The actual bit rate of TF-Codec is 6.}
\vskip 0.15in
\begin{center}
\begin{small}
\setlength{\tabcolsep}{4pt}
    \centering
    \begin{tabular}{lcccccccc}
        \toprule
        Codec & BR & SR & CER & PER & WER & SS${_{e}}$ & SS${_{s}}$ \\ \toprule
        Reference & - & - & 1.3 & 12.5 & 2.5 & 64.2 & 94.6 \\ \midrule
        Encodec & 6 & 24 & 1.7 & 12.5 & 2.7 & 56.2 & 93.2 \\
        Encodec & 12 & 24 & 1.6 & 12.4 & 2.6 & 59.1 & 93.8 \\
        TF-Codec & 8 & 16 & 1.5 & 12.4 & 2.6 & 59.4 & 94.1 \\
        DAC & 6 & 16 & 1.5 & 12.5 & 2.7 & 60.5 & 94.4 \\
        DAC & 8 & 24 & 1.4 & 12.4 & 3.1 & 61.5 & 94.4 \\
        DAC & 16 & 24 & 1.4 & 12.4 & 2.5 & 63.2 & 94.5 \\ \bottomrule
    \end{tabular}
    \label{tab:codec}
\end{small}
\end{center}
\vskip -0.1in
\end{table}

The sequential dependency of RVQ codes inspires a number of works to model them hierarchically. Since high-level and low level codes play different roles in crafting the audio, it is reasonable to separate them in multiple prediction stages. Prior works on speech generation choose to draw a hard border between these two groups of codebooks - as in AudioLM \cite{audiolm} where they separate codebooks of coarse (first 4 codebooks) and fine (remaining 8 codebooks) acoustic tokens and model them with two stacked autoregressive transformers, or in VALL-E \cite{vall-e} where they model the first codebook with an autoregressive transformer, and the rest of them with a non-autoregressive transformer. In either case, each code is predicted by a query to the transformer, and codes in a frame have to be generated sequentially in the streaming mode.

By having only one stage, MusicGen \cite{musicgen} reduces the number of queries to the transformer by $Q$ times. This is possible by shifting the codebooks so that each step predicts $Q$ codes, but only one comes from each frame. Let $\boldsymbol{C'}=\left[\bc'_1,...,\bc'_{T'}\right]$ be the shifted codes obtained from $\boldsymbol{C}$, where $T'=T+Q-1$ is the length of the shifted code sequence, then we have $\bc'_i=\left[c_{i}^{(1)}, \dots, c_{i-(Q-1)}^{(Q)}\right]$, assuming padding values for invalid code positions. Each decoding step models $p\left(\bc'_{i} \middle\vert \bc'_{< i}\right)$. The $i$-th audio frame in $\mathcal{C}$ is fully generated at the $(i+Q-1)$-th step. This provides an inexact autoregressive decomposition to model the distribution of discrete codes. Although the performance of the delayed pattern is still behind that of the flatten pattern, it produces reasonable audio quality under a limited inference budget.

\begin{figure}[t]
    \centering
    \includegraphics[width=0.4\textwidth]{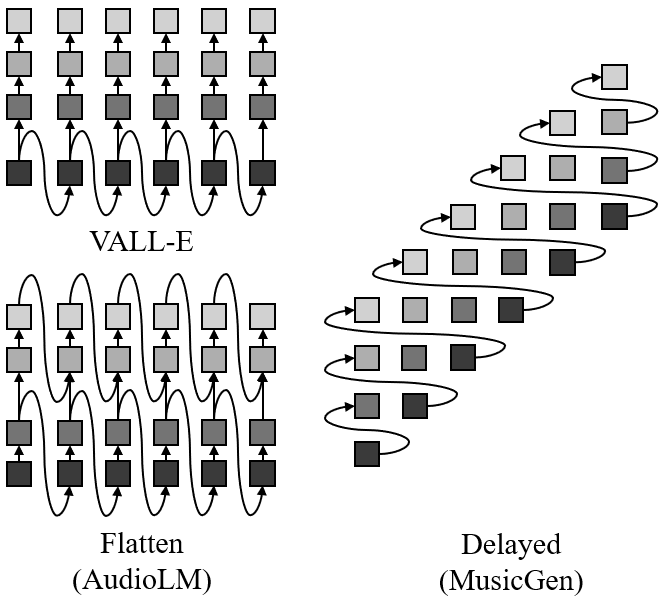}
    \caption{Different decoding pattern to generate RVQ codes with $Q=4$ codebooks: VALL-E \cite{vall-e}, Flatten \cite{audiolm}, and Delayed \cite{musicgen}. Both VALL-E and Flatten require autoregressive decoding in both the depth and width dimension, in which VALL-E uses a non-autoregressive transformer from the second codebook. The Delayed pattern only needs to perform autoregressive decoding in one dimension. Moreover, all codes in each autoregressive step can be predicted in a single transformer query.}
    \label{fig:pattern}
\end{figure}

\begin{table*}[t]  
    \caption{Decoder and Enhancer ablation results for 5s enrollment audio. Encodec is the vanilla Encodec decoder \cite{encodec}. Enhancer is DeepFilterNet3 \cite{DeepFilterNet}. Multi-Band Diffusion (MBD) \cite{MBD}}  
    \centering  
    \begin{small}  
    \setlength{\tabcolsep}{2pt}  
    \begin{tabular*}{\textwidth}{@{\extracolsep{\fill}}l c c c c c c c c c c c c}  
        \toprule  
         & \multicolumn{3}{c}{Encodec} & \multicolumn{3}{c}{Encodec + Enhancer} & \multicolumn{3}{c}{MBD} & \multicolumn{3}{c}{MBD + Enhancer} \\  
        & CER & SS & O-MOS & CER & SS & O-MOS & CER & SS & O-MOS & CER & SS & O-MOS \\ \midrule  
        Reference & 1.2 & 64.4 & 3.60 & 1.5 & 63.8 & 3.65 & 1.61 & 60.1 & 3.68 & 1.5 & 59.3 & 3.80 \\  
        MusicGen (Adapted$^{*}$) \cite{musicgen} & 11.5 & 57.2 & 3.69 & 11.7 & 56.6 & 3.74 & 12.7 & 50.0 & 3.60 & 12.9 & 48.9 & 3.82 \\  
        Ours - $\lambda=0.1$ & 3.2 & 55.7 & 3.58 & 3.6 & 54.8 & 3.67 & 4.8 & 50.5& 3.57 &  4.4 & 49.2 & 3.77 \\  
        Ours - $\lambda=0.1, p_{\max}=0.5$ & 3.8 & 56.5 & 3.57 & 3.9 & 55.5 & 3.67 & 4.4 & 50.4 & 3.59 & 4.4 & 49.3 & 3.80 \\  
        \bottomrule  
    \end{tabular*}  
    \label{tab:Decoding}  
    \end{small}  
\end{table*}  

\subsection{Token Generation Pattern}

Figure \ref{fig:pattern} compares three decoding patterns. In the delayed pattern that we use, since each query is used to predict $Q$ codes at the same time, the number of transformer steps is $Q$ times lower than other patterns.

\label{sec:appendix:pattern}

\subsection{Content Accuracy - Voice Quality Trade-off}

We train a model that prioritizes high-level codebooks by assigning higher weights to high-level codebook losses. Particularly, we initialize weights for the first 4 codebooks as 16, 8, 4, 2, and all other codebooks as 1. We apply exponential decay to these weights such that they converge to 1 at the end of training. Compared to the model without codebook loss weights, this model achieves a significantly better CER at the expense of the SS score (Table \ref{tab:tradeoff}).

\subsection{Decoding and Enhancement}
\label{sec:Decoding}

We explored utilizing a multi-band diffusion decoder (MBD) 
\cite{MBD} 
as a substitute for the Encodec decoder \cite{encodec} in order to improve audio quality. Our observations indicated that while MBD alters the speaker characteristics, resulting in a decrease in the SS score, it simultaneously improves the O-MOS score \cite{reddy2021dnsmos}, see Table \ref{tab:Decoding}. In addition, MetaVoice's findings revealed the presence of background artifacts in the decoded waveform, which prompted them to examine the application of an enhancer, such as DeepFilterNet \cite{DeepFilterNet}, to the generated waveform with the aim of removing artifacts introduced by MBD, therefore refining audio quality. We report scores when using the enhancer for both the Encodec and the MBD codec decoder in Table \ref{tab:Decoding}. A uniformly sampled subset of 148 utterances are used to report these scores.

\end{document}